\documentclass[prb,aps,twocolumn,superscriptaddress]{revtex4}
\usepackage{graphicx}
\usepackage{amsmath}
\usepackage{bm}
\begin{document}
\title{Resonant nature of phonon-induced damping of Rabi
oscillations in quantum dots}
\author{P. Machnikowski}
 \email{Pawel.Machnikowski@pwr.wroc.pl}
 \affiliation{Institute of Physics, Wroc{\l}aw University of
Technology, 50-370 Wroc{\l}aw, Poland}
 \affiliation{Institut f{\"u}r Festk{\"o}rpertheorie,
 Westf{\"a}lische Wilhelms-Universit{\"a}t, 48149 M{\"u}nster, Germany}
\author{L. Jacak}
 \affiliation{Institute of Physics, Wroc{\l}aw University of
Technology, 50-370 Wroc{\l}aw, Poland}

\begin{abstract}
Optically controlled coherent dynamics of charge (excitonic)
degrees of freedom in a semiconductor quantum dot under influence
of lattice dynamics (phonons) is discussed theoretically. We show
that the dynamics of the lattice response in the strongly
non-linear regime is governed by a semiclassical resonance between
the phonon modes and the optically driven dynamics. We stress the
importance of the stability of intermediate states for the truly
coherent control.
\end{abstract}

\maketitle

One of the most challenging proposals exploiting the atomic-like
properties of quantum dots (QDs) \cite{jacak98a,bimberg99} is to
implement the quantum information processing schemes
\cite{bouwmeester00,alber01} using confined carrier states in
these artificial systems. One of the proposed solutions is to use
the charge degrees of freedom which may be efficiently controlled
optically even on sub-picosecond scale
\cite{biolatti00,derinaldis02,chen01,piermarocchi02} which may
look promising in view of the measured 1 ns lifetime of excitons
in QDs \cite{borri01,bayer02}. The first steps towards this goal
include experimental proofs of controlled coherent dynamics in
these structures \cite{bonadeo98}, observation of Rabi
oscillations
\cite{stievater01,kamada01,htoon02,zrenner02,borri02a},
demonstration of entanglement between states of interacting dots
\cite{bayer01} and performing a quantum logic gate based on a
bi-exciton system in a QD \cite{li03}.

However, the coherent dynamics of quantum confined carrier states
is very sensitive to interaction with the macroscopic number of
degrees of freedom of the outside world. If the coherent dynamics
involves an excited state of the confined exciton
\cite{bonadeo98,kamada01,htoon02} the coherence time is limited by
relaxation to the ground state (usually $\sim 30$ ps). The
measured signal may further be affected by the particularities of
the detection technique \cite{htoon02}. Another approach consists
in measuring the dot occupation upon driving by pulses with
constant length and varying amplitudes
\cite{stievater01,zrenner02,borri02a}. So far, it has always
turned out that such pulse-area-dependent Rabi oscillations
deviate from the ideal ones, the discrepancy being larger for
stronger pulses. In principle, this might also be explained by
experimental conditions or environmental perturbation: scattering
by weakly localized excitons around an interface fluctuation QD
(further confirmed by increasing decay for stronger pulses)
\cite{stievater01}, tunneling to leads in the photodiode structure
(on $\sim 10$ ps timescale) \cite{zrenner02}, or dipole moment
distribution in the QD ensemble \cite{borri01}.

One might believe that all the perturbation comes from sources
that may be removed or minimized by technology improvement and by
optimizing the experimental conditions and hence produce no
fundamental obstacle to arbitrarily perfect quantum control over
the excitonic states. However, in every case the QDs are
inherently coupled to the surrounding crystal lattice. The
perturbing effect of lattice modes (phonons) has been observed
experimentally as a fast ($\sim 1$ ps) partial decay of optical
polarization induced by an ultra-fast laser pulse \cite{borri01}.
The theoretical analysis \cite{krummheuer02,vagov02a,jacak03b}
shows that the experimentally observed effects may be
quantitatively accounted for by invoking the carrier-phonon
coupling. Indeed, under suitable excitation conditions in a
high-quality sample the carrier density may be kept low enough to
eliminate Coulomb scattering effects while the coupling to the
electromagnetic modes is manifested by the exciton radiative decay
on timescales many order of magnitude longer than these relevant
here. The theory shows that the coherence decay should be viewed
as a trace of coherent lattice dynamics (due to lattice inertia),
persisting even at zero temperature when the lattice is initially
at its ground state and cannot perturb the stationary ground state
of the carrier subsystem. Due to strong reservoir memory on
timescales relevant for these processes, they cannot be fully
understood within the Markovian approximations. In particular, the
idea of a ``decoherence time'', with which the control dynamics
competes, is misleading \cite{alicki02a}.

The recent theoretical study \cite{forstner03} on optical Rabi
flopping of excitons in QDs driven by finite-length optical pulses
shows that exponential damping models fail to correctly describe
the system kinetics. The appropriate description yields much less
damping, especially for long pulses. It turns out that the lowest
``quality'' of the Rabi oscillation is obtained for pulse
durations of a few ps, while for longer durations the damping is
again decreased.

In this paper we propose both qualitative and quantitative
explanation of the mechanism leading to the phonon-induced damping
reported in the theoretical and experimental studies. We show that
the carrier-phonon interaction responsible for the damping of the
oscillations has a resonant character: While in the linear limit
the system response depends only on the spectral decomposition of
the pulse, the situation is different when a strong pulse induces
an oscillating charge distribution in the system. In a
semi-classical picture, this would act as a driving force for the
lattice dynamics. If the induced carrier dynamics is much faster
than phonon oscillations the lattice has no time to react until
the optical excitation is done. The subsequent dynamics will lead
to exciton dressing, accompanied by emission of phonon packets,
and will partly destroy coherence of superposition states
\cite{borri01,krummheuer02,vagov02a,jacak03b} but cannot change
the exciton occupation number. In the opposite limit, the carrier
dynamics is slow enough for the lattice to follow adiabatically.
The optical excitation may then be stopped at any stage without
any lattice relaxation incurred, hence with no coherence loss. The
intermediate case corresponds to modifying the charge distribution
in the QD with frequencies resonant with the lattice modes which
leads to increased interaction with phonons and to decrease of the
carrier coherence (see Ref. \onlinecite{axt99} for a simple,
single-mode model).

The above ideas are supported by the formal analysis in the
following. First, we describe the theoretical method used; then
the description is applied to a specific self-assembled InAs/GaAs
quantum dot to get quantitative estimations; finally, we conclude
with some summary and possible consequences of the results.

Following the earlier studies \cite{krummheuer02,vagov02a,vagov03}
successfully accounting for the observed carrier dynamics on
picosecond timescales, we describe the system by the RWA
Hamiltonian (in the rotating frame)
\begin{displaymath}
H=H_{\mathrm{X}}+H_{\mathrm{ph}}+H_{\mathrm{int}}.
\end{displaymath}
Here $H_{\mathrm{X}}=f(t)(a+a^{\dagger})$ describes the
interaction of the ground-state exciton with the resonant laser
pulse modulated by the real envelope function $f(t)$
($a,a^{\dagger}$ are excitonic operators),
$H_{\mathrm{ph}}=\sum_{\bm{k}}\omega_{\bm{k}}
b^{\dagger}_{\bm{k}}b_{\bm{k}} $ is the lattice Hamiltonian
($b^{\dagger}_{\bm{k}},b_{\bm{k}}$ refer to the phonon mode
$\bm{k}$ whose energy is $\omega_{\bm{k}}$), and
\begin{displaymath}
H_{\mathrm{int}}=a^{\dagger}a \sum_{\bm{k},s}\left(
f_{\bm{k}}b^{\dagger}_{\bm{k}}+f^{*}_{\bm{k}}b_{\bm{k}} \right)
\end{displaymath}
is the interaction Hamiltonian ($f_{\bm{k}}$ are the coupling
constants for the individual phonon modes). Due to momentum
conservation, only long wavelength phonon modes (compared to dot
size) are effectively coupled \cite{jacak02a,jacak03a}. For the
timescales relevant here, the damping effect is caused mostly by
acoustic phonons \cite{vagov03, forstner03}. In the calculations,
we include deformation potential coupling to longitudinal
acoustical (LA) phonons, which describes the modification of the
band structure due to lattice compression and is different for
electrons and holes. Since no cancellation is involved, its effect
depends only weakly on the exact wavefunction geometry. We assume
that transitions to higher states, including possible inter-valley
transitions, are well separated in energy, compared to the
spectral characteristics of the dynamics considered here and may
be neglected. Also bi-exciton transitions are assumed to be
eliminated by a suitable polarization choice. Piezoelectric
coupling to acoustic phonons has been shown to be negligible if
the electron and hole function overlap\cite{krummheuer02} and is
neglected here (but might contribute if the  wavefunction geometry
is different). We have verified that the Fr{\"o}hlich coupling to
the longitudinal optical (LO) phonons is negligible, due not only
to charge cancelation but mostly to large LO phonon frequencies,
compared to the timescales involved.

Initially, the state of the system is described by the density matrix
$\varrho=\varrho_{0}\otimes\varrho_{\mathrm{l}}(T)$,
where $\varrho_{0}$ is the state with no exciton and
$\varrho_{\mathrm{l}}(T)$ is the thermal equilibrium of lattice modes at
the temperature $T$.
The dynamics of the system is calculated by the perturbation expansion of
the evolution operator in Born approximation.
The
lattice degrees of freedom are then traced out leading to the reduced
density matrix for the carrier subsystem, which contains all the
information accessible by optical methods. The details of the procedure
may be found in \cite{alicki02a}. As a result, the reduced density matrix
after time $t$ may be written as
$ \varrho(t)=U(t)(\varrho_{0}+\varrho_{1} )U^{\dagger}(t)$,
where $U(t)$ is the driven evolution for the unperturbed carrier system
and $\varrho_{1}$ describes the perturbation to the carrier dynamics
coming from phonons. The matrix elements of the latter, in the basis of
empty dot ($|0\rangle$) and one exciton ($|1\rangle$) states,
may be written as
\begin{subequations}
\begin{eqnarray}
\langle 0 | \varrho_{1} | 0\rangle
    = -\langle 1 | \varrho_{1} | 1\rangle & = &
-\int_{-\infty}^{\infty}d\omega
    \frac{R(\omega)}{\omega^{2}}S_{00}(\omega),
\label{ro00} \\
\langle 0 | \varrho_{1} | 1\rangle
    = \langle 1 | \varrho_{1} | 0\rangle^{*}
& = &
-\int_{-\infty}^{\infty}d\omega\
        \frac{R(\omega)}{\omega^{2}}S_{01}(\omega).
\label{ro01}
\end{eqnarray}
\end{subequations}
An additional component describing a unitary correction, e.g.
light coupling renormalization and energy shifts, may be canceled
by an appropriate modification of $H_{\mathrm{X}}$ (these effects
may lead e.g. to intensity dependence of the observed Rabi
frequency\cite{forstner03}). The function $R(\omega)$ is the
spectral density of the reservoir, fully characterizing the
lattice properties at this order of perturbation treatment,
defined as
\begin{displaymath}
R(\omega) = \sum_{\bm{k}}|f_{\bm{k}}|^{2} \left[
(n_{\bm{k}}+1)\delta(\omega-\omega_{\bm{k}})
+n_{\bm{k}}\delta(\omega+\omega_{\bm{k}}) \right].
\end{displaymath}
It depends on the material parameters and QD size via the coupling
constants $f_{\bm{k}}$ (see e.g. \cite{jacak03b} for explicit formulas).
The functions $S_{00}(\omega),S_{01}(\omega)$ contain the complete
information on the optically controlled carrier dynamics.
If we assume that final state is measured after a time long compared to
the phonon oscillation periods, they may be written as
\begin{eqnarray*}
S_{00}(\omega) & = &
\frac{1}{4}\left[ \sin^{2}\alpha+|K_{\mathrm{s}}(\omega)|^{2} \right], \\
S_{01}(\omega) & = & \frac{1}{8}\left[ \sin 2\alpha
+2\mathrm{Re}K_{\mathrm{s}}(\omega) K^{*}_{\mathrm{c}}(\omega)
\right],
\end{eqnarray*}
where
\begin{subequations}
\begin{eqnarray}
K_{\mathrm{s}}(\omega) & = & \int_{s}^{t}d\tau e^{i\omega\tau}
\frac{d}{d\tau}\sin 2F(\tau), \label{ks} \\
K_{\mathrm{c}}(\omega) & = & \int_{s}^{t}d\tau e^{i\omega\tau}
\frac{d}{d\tau}\cos 2F(\tau), \label{kc}
\end{eqnarray}
\end{subequations}
where $s,t$ are the initial and final time of the evolution,
$ F(t)=\int_{s}^{t}d\tau f(\tau)$
is the rotation (on the Bloch sphere) performed on the exciton state up
to time $t$ and $\alpha$ is the total angle of rotation.
For a given $\alpha$, the functions
(\ref{ks},b) actually depend only on $\omega\tau_{\mathrm{p}}$, where
$\tau_{\mathrm{p}}$ is the pulse duration.

\begin{figure}
\unitlength 1mm
\begin{center}
\begin{picture}(80,45)(0,8)
\put(0,0){\resizebox{80mm}{!}{\includegraphics{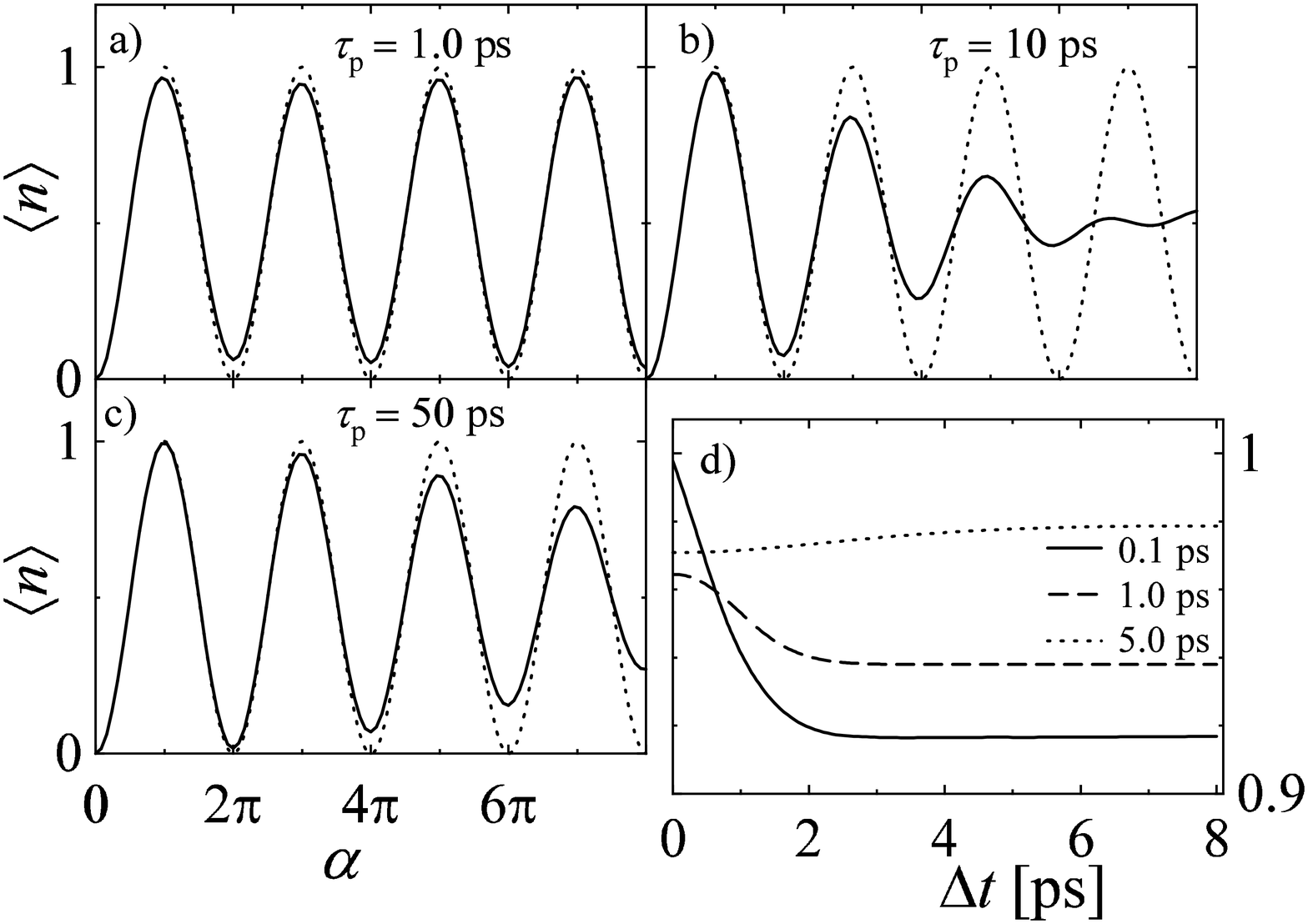}}}
\end{picture}
\end{center}
\caption{\label{fig:rabi}(a-c) Pulse-area-dependent Rabi
oscillations for various pulse durations $\tau_{\mathrm{p}}$ as
shown in the figure, for $T=10$ K ($\alpha$ is the rotation angle
on the Bloch sphere). Dotted line shows unperturbed oscillations.
(d) The final QD occupation after two $\pi/2$ pulses separated by
time interval $\Delta t$ for pulse durations as shown.}
\end{figure}

In a typical experiment
\cite{stievater01,zrenner02,borri02a} one measures the average dot
occupation, $\langle n\rangle=\langle 1|\varrho(\infty)|1\rangle$,
after a pulse of fixed length but variable amplitude. The
simulation of such an experiment within the presented model is
shown in Fig. \ref{fig:rabi} for Gaussian pulses
[$\tau_{\mathrm{p}}$ is the full width at half maximum of the
pulse envelope $f(t)$]. On the grounds of earlier numerical
calculations \cite{jacak03b}, we have assumed the exciton ground
state to be of nearly product character, with Gaussian electron
and hole wavefunctions with in-plane localization widths of 4.9 nm
and 4.0 nm, respectively and with growth-direction width of 1 nm
(this corresponds to a dot of roughly 20 nm diameter and 4 nm
height). The oscillations are almost perfect for very short pulses
($\sim 1$ ps), then loose their quality for longer
pulse durations ($\sim 10$ ps). Although this might be expected
from any simple decoherence model, the striking feature is that
the effect dramatically grows for higher oscillations, despite the
fact that the whole process has exactly constant duration. Even
more surprising is the improvement of the quality of oscillations
for long pulses ($\sim 50$ nm) where, in addition, the first
oscillation is nearly perfect.

\begin{figure}
\unitlength 1mm
\begin{center}
\begin{picture}(80,45)(0,8)
\put(0,0){\resizebox{80mm}{!}{\includegraphics{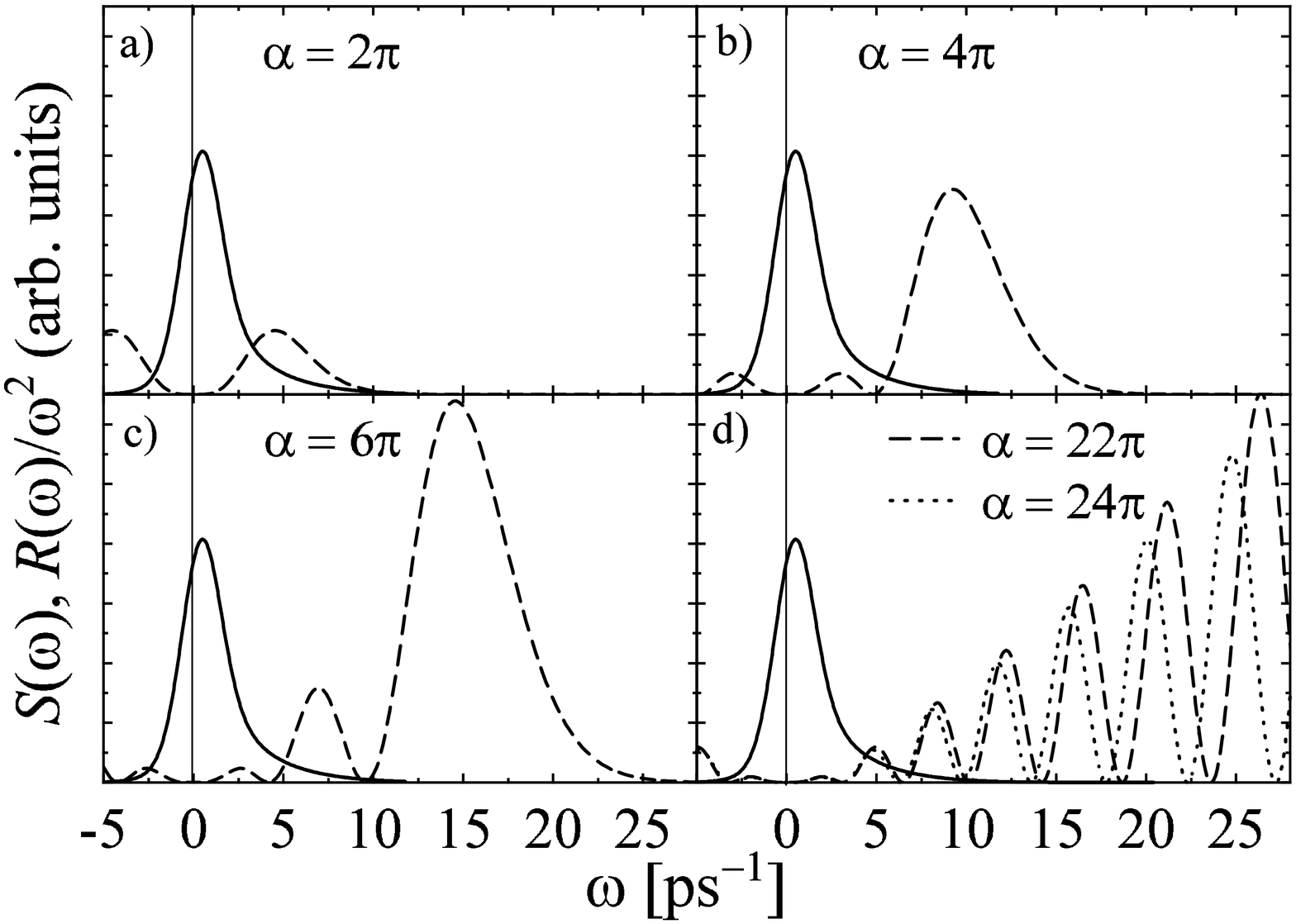}}}
\end{picture}
\end{center}
\caption{\label{fig:sr}Phonon spectral density
$R(\omega)/\omega^{2}$ (solid) for deformation potential coupling
to LA phonons at $T=10$ K and the nonlinear pulse spectrum
$S_{00}(\omega)$ (dashed and dotted lines) for
$\tau_{\mathrm{p}}=1$ ps, for rotation angles $\alpha$ as shown.}
\end{figure}

The formulas (\ref{ro00},b) quantify the idea of resonance between
the induced dynamics and lattice modes: in Fig. \ref{fig:sr} the
phonon spectral density is compared to the non-linear frequency
characteristics of the optically controlled exciton dynamics for
$\tau_{\mathrm{p}}=1$ ps (the characteristics for other durations
is easily obtained by scaling). According to (\ref{ro00},b), the
overlap of these spectral characteristics with the phonon spectral
density gives the perturbation of the coherent carrier dynamics.

For growing number of rotations $n$, the nonlinear pulse spectrum
$S_{00}(\omega)$ develops a series of maxima of increasing
strength (Fig. \ref{fig:sr}a,b,c). The position of the last and
highest maximum corresponds approximately to $2\pi
n/\tau_{\mathrm{p}}$, in accordance with the semiclassical
resonance concept. However, spectral components are also present
at all the frequencies $2\pi n'/\tau_{\mathrm{p}}$, $n'<n$, which
is due to the turning on/off of the pulse. It is interesting to
note that for high $n$, the low-frequency part of $S_{00}(\omega)$
does not evolve with $n$ (Fig. \ref{fig:sr}d). It is now clear
that there are two ways of minimizing the overlap: either the
pulse must be so short that all the maxima of $S_{00}(\omega)$ are
pushed to the right into the exponentially vanishing tail of the
reservoir spectral density $R(\omega)$, or the pulse must be very
long, to ``squeeze'' the spectral function near $\omega=0$ and
thus reduce its area. In the latter case, the maxima developing
with growing number of oscillations will eventually overlap with
$R(\omega)$ destroying the coherence.

Although it might seem that speeding up the process is the
preferred solution, it is clear that this works only because no
high frequency features are included into the present model. In
reality, speeding up the dynamics is limited e.g. by the presence
of excited states and non-adiabatically enhanced LO phonon
coupling \cite{fomin98}. Moreover, it turns out that the resulting
dynamics, even within this model, is actually not fully coherent.
It has been shown \cite{vagov02a} that superposition of states
created by an ultra-short $\pi/2$ pulse becomes corrupted,
preventing a second pulse (after some delay time $\Delta t$) from
generating the final state of $\langle n\rangle=1$ with unit
efficiency. In order to prove the fully coherent character of
carrier dynamics it is necessary to demonstrate the stability of
the intermediate state in a two-pulse experiment. The simulations
of such an experiment are shown in Fig. \ref{fig:rabi}d. A short
pulse ($\tau_{\mathrm{p}}=0.1$ ps) creates a superposition of bare
states (surrounded by non-distorted lattice) which then decohere
due to dressing processes \cite{krummheuer02,jacak03a}. As a
result, the exciton cannot be created by the second pulse with
unit probability \cite{vagov02a}. For a longer pulse
($\tau_{\mathrm{p}}=1.0$ ps), the lattice partly manages to follow
the evolution of charge distribution during the optical operation
and the destructive effect is smaller. Finally, if the carrier
dynamics is slow compared to the lattice response times
($\tau_{\mathrm{p}}\sim 10.0$ ps), the lattice distortion follows
adiabatically the changes in the charge distribution and the
superposition created by the first pulse is an eigenstate of the
interacting carrier-lattice system, hence does not undergo any
decoherence and the final effect is the same for any delay time
(its quality limited by decoherence effects during pulsing). In
fact, splitting the $\pi$--pulse into two corresponds to slowing
down the carrier dynamics which, in the absence of decoherence
during delay time, improves the quality of the final state, as
seen in Fig \ref{fig:rabi}d.

The above analysis shows that damping of pulse-area-dependent Rabi
oscillations due to interaction with lattice modes is a
fundamental effect of non-Markovian character: it is due to a
semiclassical resonance between optically induced confined charge
dynamics and lattice modes. The destructive effect may be
minimized both by speeding up and slowing down the dynamics.
However, in the former case, the system passes through unstable
(decohering) states. Moreover, fast operation on a real system
induces many undesirable effects: transitions to higher states,
bi-exciton generation or resonant LO phonon dynamics. On the other
hand, for slow operation, the number of ``good'' oscillations is
limited. Thus, it is impossible to perform an arbitrary number of
fully coherent Rabi oscillations on an exciton confined in a
quantum dot.

By increasing the pulse duration as
$\tau_{\mathrm{p}}\sim\alpha^{2}$, the phonon effect on the
exciton dynamics may be kept constant. However, in this case the
achievable number of oscillations is strongly restricted by the
exciton lifetime and other (thermally activated) processes.
Eliminating the radiative losses e.g. by using stimulated Raman
adiabatic passage instead of a simple optical excitation
\cite{troiani03} seems to be a promising direction from this point
of view.

The model presented above accounts for the decrease of the quality of
Rabi oscillations in the short duration range observed in the experiment
\cite{borri02a}. It predicts, however, that this trend is reversed
for longer pulse durations. The quantitative value of 96\% for the first
maximum of the oscillations with $\tau_{\mathrm{p}}=1$ ps agrees very
well with the experimental result \cite{zrenner02} although the following
extrema are much worse in reality than predicted here.
This suggests an increased
lattice response at higher frequencies which may be due to more
complicated wavefunction geometry
\cite{heitz99} or electric-field induced charge
separation leading to strong piezoelectric effects \cite{krummheuer02}.
Nevertheless, as far as it may be inferred from the experiment
\cite{zrenner02}, the decrease of the oscillation quality seems to
saturate after one full Rabi rotation, as predicted by the model
calculations.

In conclusion, we have studied the carrier-lattice dynamics for
optically induced Rabi oscillations of exciton occupation in a
quantum dot. We have shown that the lattice response is resonantly
driven by a combination of the (linear) pulse spectrum and the
Rabi frequency. This semiclassical interpretation allows one to
qualitatively predict the damping effect based on the general
knowledge of the spectral properties of the lattice modes and
leads to the conclusion that the resonant lattice response
precludes performing a large number of fully coherent oscillations
in this system.

We are grateful to T. Kuhn for reading the manuscript and many
useful comments. P.M. is grateful to M. Horodecki for discussions
and to the Alexander von Humboldt Foundation for generous support.
Supported by the Polish Ministry of Scientific Research and
Information Technology under the (solicited) grant No
PBZ-MIN-008/P03/2003 and by the Polish KBN under grant No. PB 2
PO3B 085 25.


\end{document}